# Understanding the chemical mechanism behind photo-induced enhanced Raman spectroscopy

Junzhi Ye[#,1,6], Rakesh Arul[#*,1,2,3,4,6], Michel K. Nieuwoudt[1,2,3,4], Junzhe Dong[6], Wei Gao*[6], M. Cather Simpson*[1,2,3,4,5]

1. The Photon Factory, the University of Auckland, Auckland, New Zealand
2. The MacDiarmid Institute for Advanced Materials and Nanotechnology, New Zealand
3. The Dodd Walls Centre for Quantum and Photonic Technologies, New Zealand
4. School of Chemical Sciences, the University of Auckland, Auckland, New Zealand
5. Department of Physics, the University of Auckland, Auckland, New Zealand
6. Department of Chemical and Materials Engineering, the University of Auckland, Auckland, New Zealand

* c.simpson@auckland.ac.nz, w.gao@auckland.ac.nz, ra554@cam.ac.uk, [#] These authors contributed equally to the work

## Abstract

Photo-Induced Enhanced Raman Spectroscopy (PIERS) is a new surface enhanced Raman spectroscopy (SERS) modality with an order-of-magnitude Raman signal enhancement of adsorbed analytes over that of typical SERS substrates. Despite the impressive PIERS enhancement factors and explosion in recent demonstrations of its utility, the detailed enhancement mechanism remains undetermined. Using a range of optical and X-ray spectroscopies, supported by density functional theory calculations, we elucidate the chemical and atomic-scale mechanism behind the PIERS enhancement. Stable PIERS substrates with enhancement factors of $10^6$ were fabricated using self-organized hexagonal arrays of $TiO_2$ nanotubes that were defect-engineered via annealing in inert atmospheres, and silver nanoparticles were deposited by magnetron sputtering and subsequent thermal dewetting. We identified the key source of the enhancement of PIERS vs. SERS in these structures as an increase in the Raman polarizability of the adsorbed probe molecule upon photo-induced charge transfer. A balance between crystallinity, which enhances charge transfer due to higher electron mobility in anatase-rutile heterostructures but decreases visible light absorption, and oxygen vacancy defects, which increase visible light absorption and photo-induced electron transfers, was critical to achieve high PIERS enhancements.

## Introduction

Raman spectroscopy is a widely used tool in the analytical sciences, enabling highly sensitive fingerprinting of molecules for healthcare, security, and environmental applications. However,

Raman scattering is an inefficient process, and only accounts for one in ten million scattered photons[1]. Surface enhanced Raman spectroscopy (SERS) uses nanostructured plasmonic metals (e.g. Au, Ag, Al) supporting a localized surface plasmon resonance that enhances the intensity of incident laser light in the near-field. The enhancement of Raman scattering intensities can be spectacular, with state-of-the-art techniques (combining SERS with resonance Raman processes) reaching enhancement factors values as large as $10^9$ [2], with typical values between $10^4$-$10^8$ [3]. Recently, the use of dielectrics as support or active materials in SERS has grown. Many dielectrics can act as passive elements supporting plasmonic metal nanostructures to localize electromagnetic fields via a micro-lensing effect [4], and providing an inert shell that assists in-situ SERS measurements (e.g. SHINERS [5] or SPARKs [6]). Semiconductors may also act as active SERS substrates via charge transfer to the target analyte molecule [7], and aid reusability via photocatalytic degradation of adsorbed molecules upon exposure to UV light. However, these techniques have not achieved comparable SERS enhancements (only $10^2$-$10^5$ [8]), or work as reproducibly as metal nanoparticle-based SERS.

A new development that aims to push the enhancement factor of Raman beyond that of SERS is Photo-Induced Enhanced Raman Spectroscopy (PIERS) [9]. In PIERS, a defect engineered semiconductor (e.g. $TiO_2$) acts in concert with plasmonic nanoparticles to enhance the Raman scattering by an order of magnitude relative to SERS. The hypothesis for the enhancement mechanism described in the original discovery by Ben Jaber et. al. [9a] is as follows: $TiO_2$ containing oxygen vacancies can absorb visible light due to defect levels in the conduction band. Once in the conduction band, mobile electrons migrate to the plasmonic nanoparticle, and charge transfer occurs from the metal to the adsorbed molecules. Prior works have proposed a charge-transfer based enhancement for many dye molecules in pure/defect engineered semiconductors[10], semiconductor heterostructures[11] or semiconductor-metal heterostructures [12]. However, such works have used conditions in which the dye molecule is on- or near-resonance with the excitation laser[10a, 10c, 10f, 11b, 13], thus having resonance Raman enhancements in addition to substrate-induced enhancements, or claim charge transfer without providing evidence for the exact mechanism [10d, 10e, 14]. Indeed only a few studies [10a, 10g, 12, 13c, 15] show evidence that bona-fide charge transfer resonances between localized molecular states to semiconductor bands are boosting the observed Raman signals, while others [16] suggest this but do not provide any direct evidence of thermodynamically favourable band alignments. No effort has been made yet to map out the energy levels involved to establish the mechanism.

In this work we map out the mechanism of the PIERS process. We found that instead of the commonly accepted charge transfer resonance Raman enhancement model proposed [17] [9a, 10h], the most likely mechanism of enhancement is one that involves the increase of the intrinsic polarizability of an adsorbed molecule upon charge transfer. We also demonstrate a new method for making stable, reproducible PIERS substrates that utilizes simple annealing, sputtering and self-organization techniques.

# Results and Discussion

## Enhanced Raman signals in the PIERS v.s. SERS substrates

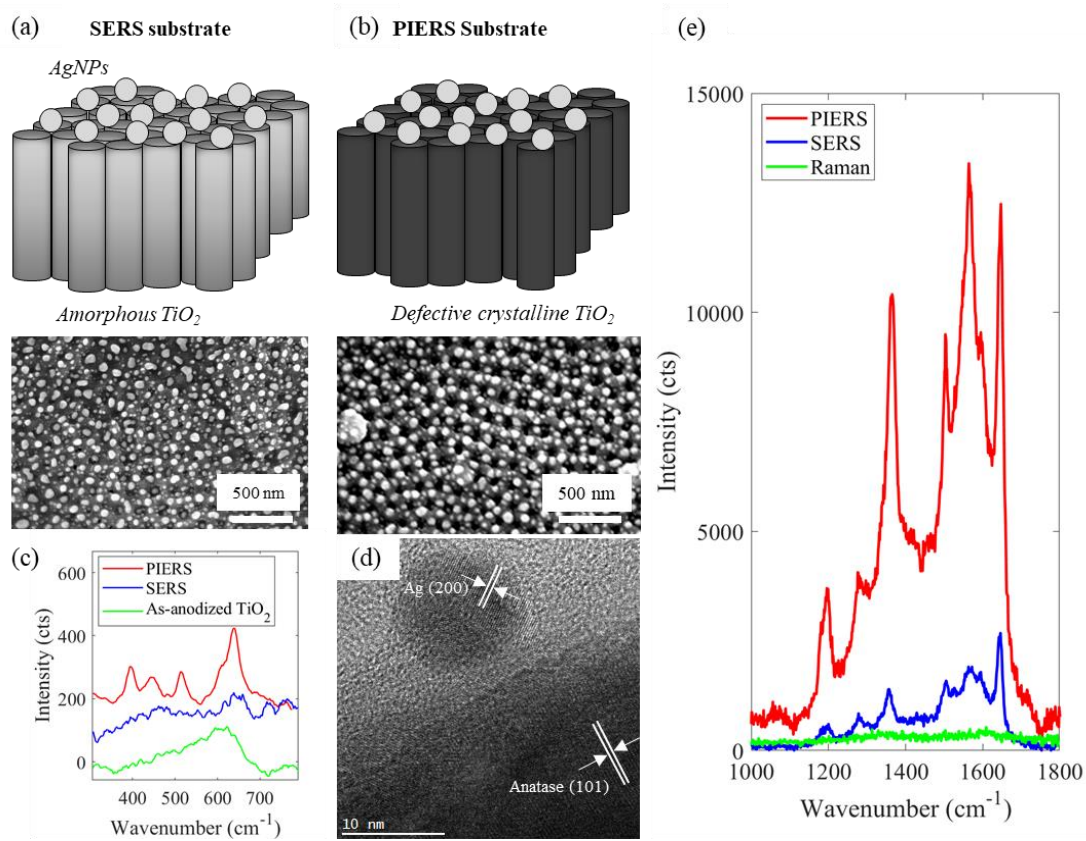

Figure 1. PIERS vs SERS substrate characteristics and enhanced Raman signals of PIERS vs SERS. (a) SEM image of SERS (Ag nanoparticles, AgNPs on amorphous $TiO_2$) substrate and (b) PIERS (AgNPs on crystalline defective $TiO_2$ - annealed at 600°C for 1 hour in Ar prior to AgNP formation) substrate with schematics of SERS and PIERS substrates above. (c) Raman spectrum of Rhodamine B ($10^{-5}$M) on a bare $TiO_2$ substrate, a SERS substrate (enhancement factor $2\times10^5$), and a PIERS substrate (enhancement factor $1\times10^6$). (d) TEM image of PIERS substrate with lattice fringes and phase indicated (further detail in Fig. S7). (e) Raman spectra of bare as-anodized $TiO_2$ substrate, SERS substrate, and PIERS substrate showing crystalline anatase and rutile Raman bands.

SERS and PIERS substrates were synthesized by thermally dewetting silver nanoparticles (AgNPs) on the surface of anodized $TiO_2$ nanotube arrays. We controlled the crystallinity of $TiO_2$ by thermal annealing. The $TiO_2$ annealing temperature, annealing environment (air v.s. argon), Ag dewetting temperature, and the thickness of the magnetron-sputtered Ag layer were varied to achieve a large enhancement factor, suitable plasmon resonance, and high degree of uniformity over the surface (SI Table S2). The morphology of the resulting substrates is shown in Fig 1(a) and (b). The AgNPs were uniformly distributed on the surface of the $TiO_2$ nanotube array and were well dispersed around the rims of the nanotubes. The AgNPs exhibited a bimodal size distribution clustering around 5-15 nm and 65-75 nm (particle size distribution histogram in SI Fig. S4). The particle density of resonant AgNPs was consistent between the SERS and PIERS substrates, and many SERS hotspots lie underneath the 1-2 µm diameter Raman laser probe.

The PIERS substrate's surface atomic structure is a mixture of crystalline anatase and rutile, as shown by anatase-specific Raman peaks at 395 ($B_{1g}$), 515 ($A_{1g}$), and 639 ($E_g$) $cm^{-1}$ [18] and rutile-specific peaks at 446 ($E_g$) and 606 ($A_{1g}$) $cm^{-1}$ in Fig 1(c). The SERS and as-anodized $TiO_2$ substrates are amorphous (Fig 1(c)). The characterization of the PIERS substrate's structure is supported by high resolution TEM images (Fig 1(d)) that show AgNPs in close contact with the anatase and rutile phases (Fig S7). X-ray diffraction measurements (Fig S8) enable us to determine a 35% rutile to 65% anatase phase fraction for the PIERS substrate.

The PIERS effect is clearly visible in Fig 1(e), in which the PIERS Raman spectra of adsorbed Rhodamine B (RhB) display an order-of-magnitude greater enhancement factor over the equivalent SERS substrates. This is similar to Ben-Jaber et al.'s original observation [9a]. Fig. 2 depicts the spatially averaged enhanced Raman spectra (over 10 different locations), using Rhodamine B (RhB) as a probe molecule. The enhancement factors listed in Table 1 below are averaged over the surface of the substrate and calculated using several different peaks. Detailed calculations of the enhancement factor, and evaluations of the substrate's reproducibility can be found in the supplementary information (SI Section I and II). In Fig. S6, we demonstrate the effect of the AgNP size distribution by tuning the Ag dewetting time to sweep its surface plasmon into and out of resonance with the 488 nm Raman excitation laser. A roughly order of magnitude enhancement in PIERS relative to SERS is always observed, despite changing the particle size distribution, indicating this result is robust to AgNP distribution. Clearly, the PIERS effect exists for a range of different AgNP sizes.

Table 1 Enhancement Factors for SERS v.s. PIERS substrates

| Substrate | Concentration of RhB [M] | Enhancement Factors | | |
|---|---|---|---|---|
| | | *1648 $cm^{-1}$* | *1358 $cm^{-1}$* | *Average* |
| **PIERS** | $10^{-5}$ | $8.3 \times 10^5$ | $1.4 \times 10^6$ | $1.1 \times 10^6$ |
| **SERS** | $10^{-5}$ | $1.8 \times 10^5$ | $1.9 \times 10^5$ | $1.9 \times 10^5$ |

# Mechanism of the PIERS effect

Below, we map out key steps (Fig 2) in the flow of electrons induced by the Raman laser to uncover the source of the PIERS enhancement and its difference to conventional SERS.

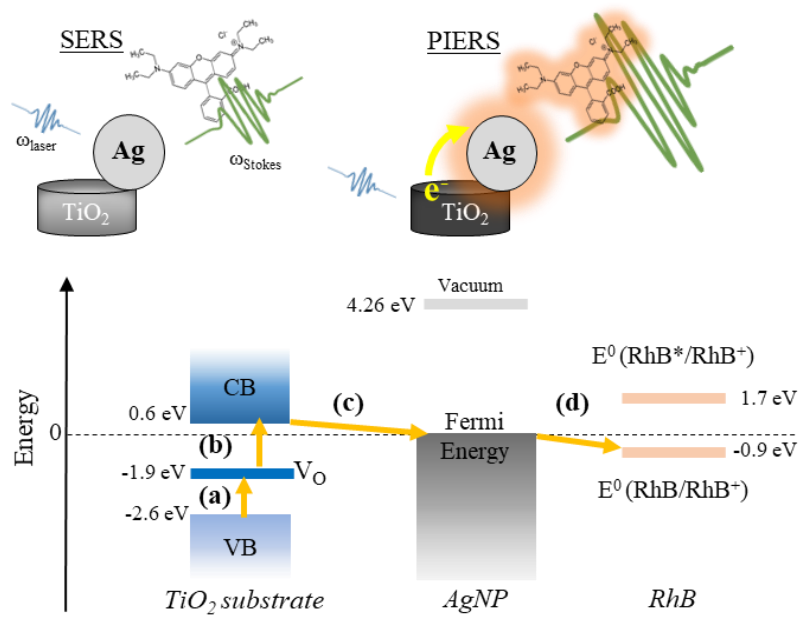

*Figure 2. Photo-Induced Enhanced Raman Spectroscopy (PIERS) mechanism: (a) Excitation of electrons in $TiO_2$ from the valence band to the defective oxygen vacancy states, (b) Excitation of electrons from defect states to the conduction band, (c) Electron transfer from $TiO_2$ to AgNPs, and (d) Electromagnetic enhancement and charge transfer interaction of AgNPs with the RhB probe molecule.*

## (a) Excitation of electrons in $TiO_2$ from the valence band to defect states

The first step in our proposed mechanism (Fig 2(a)) involves the absorption of the 488 nm laser light by oxygen vacancy defect states in within the PIERS substrate. In the corresponding SERS and as-anodized substrates, amorphous $TiO_2$ has low visible light absorption due to its broad electronic band gap [19]. UV-Vis spectra of the as-anodized sample (Fig. 3(a)) shows the typical band-edge of amorphous $TiO_2$ modulated by a thin-film interference resonance at ~350 nm. Annealing at 600°C in argon transforms amorphous $TiO_2$ to defective crystalline (anatase+rutile) $TiO_2$ (Fig 1(d)). After annealing, the substrate's color changed from yellow to black (Fig. S11), with an increased absorbance at wavelengths >400 nm (Fig. 3(a)) [19a]. The energy level of the oxygen vacancy states was measured by fitting the absorption spectrum to a Tauc plot (Fig. S12), and found to be 0.7 eV from the valence band of $TiO_2$. After the deposition and thermal de-wetting of AgNPs, Ag surface plasmon resonances were visible at 443 nm and 400 nm in the PIERS and SERS substrates (Fig. 3(a)). The broad plasmon peaks are nearly resonant with the incident 488 nm Raman laser. The Ti 2p XPS spectrum (Fig. 3(b)) of the PIERS substrate also showed the presence of $Ti^{3+}$ peaks, associated with oxygen vacancies, in the asymmetric tail at 456.9 and 462.5 eV in addition to the usual $Ti^{4+}$ peaks at

458.7 and 464.4 eV [20]. Both the XPS and UV-Vis spectra point to the presence of oxygen vacancies in PIERS substrates.

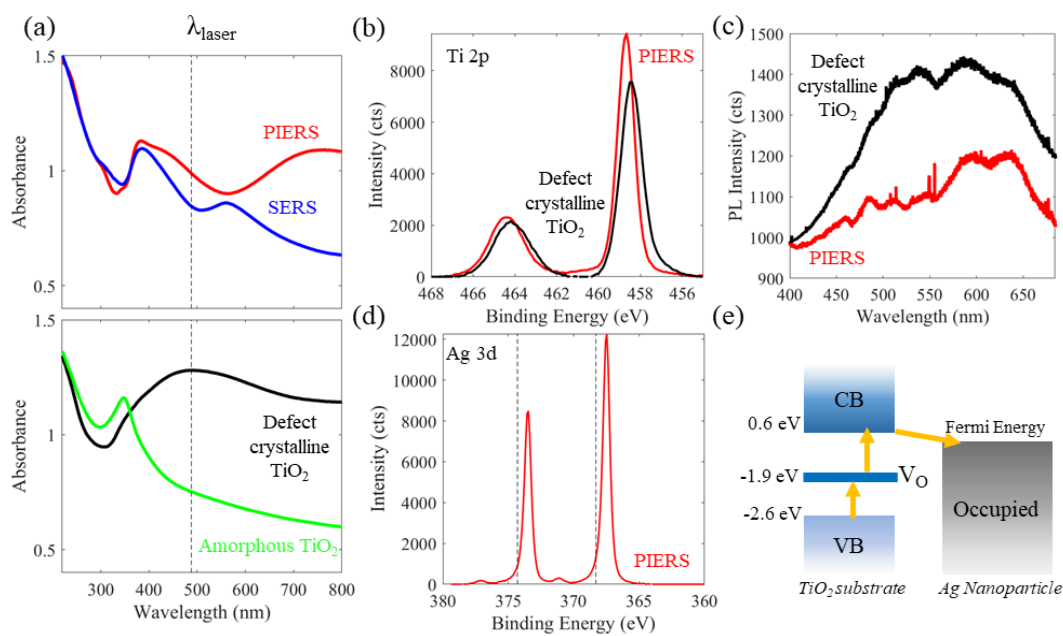

*Figure 3. Characterization of oxygen vacancies and charge transfer from TiO$_2$ to AgNPs; (a) Diffuse reflectance UV-Vis spectra of top: PIERS and SERS substrates; bottom: Defect crystalline TiO$_2$ (annealed at 600°C in Ar for 1 hour) and amorphous TiO$_2$ samples. The grey dotted line indicates the center wavelength of the Raman excitation laser (488 nm). (b) Ti 2p XPS of the PIERS and SERS substrates. (c) Photoluminescence spectra of PIERS and SERS substrate (325 nm excitation laser, 1 mW), (d) Ag 3d XPS of PIERS substrate, the vertical dotted lines are the standard binding energies of metallic Ag [21]. (e) Schematic of the initial charge transfer steps suggested by the spectroscopic measurements.*

**(b) Excitation of electrons from defect states to the conduction band**

Both the SERS and PIERS substrates have defect states due to the presence of carbon and fluorine impurities (Fig. S9) [22], in addition to oxygen vacancy defect states. This was measured in the Tauc plot (Fig. S12) which showed a reduced onset of band absorption for both SERS and PIERS substrates (2.2 and 2.4 eV respectively) compared to that for pure anatase (3.2 eV). However, strong absorption of light by defect states is a necessary but not sufficient condition to facilitate the charge transfer processes required for PIERS enhancement. The crystalline PIERS substrate (Fig 1(d)&(e)) allows electrons excited from defect states into TiO$_2$'s conduction band to propagate the distance necessary to transfer into the nearest AgNP. SERS substrates however, are amorphous and have increased radiative and non-radiative recombination, which decreases the density of photo-excited electrons that can undergo charge transfer into AgNPs. The crystallinity of the PIERS substrate is confirmed using Raman spectroscopy, X-ray diffraction, and TEM measurements in Fig. 1.

## (c) Electron transfer from TiO$_2$ to Ag

Photoinduced electron transfer from TiO$_2$ to Ag on the PIERS substrate was observed using photoluminescence (PL) spectroscopy with a 325 nm excitation laser (Fig. 3(c)). The PL spectra's peaks can be mainly assigned to self-trapped excitons (516 nm) [23], hydroxyl surface states (463 & 591 nm) [24], and recombination at oxygen vacancy states (546 & 632 nm) [25]. Fig. 3(c) shows a significant suppression in PL peak intensity in the PIERS substrate (with AgNPs) compared to the defective crystalline TiO$_2$ substrate (annealed at 600°C for 1 hour without AgNPs). The PL intensity reduction is due to the reduced radiative recombination of electrons in TiO$_2$ due to the electrons being transferred to AgNPs, with backward transfer prohibited due to the Schottky barrier formed between Ag-TiO$_2$ [26]. In addition to the photo-induced charge transfer, there is strong electrostatic Ag-TiO$_2$ charge transfer. The Ti$^{4+}$ 2p$_{3/2}$ peak increases from 458.4 eV in the defect TiO$_2$ substrate to 458.7 eV in the PIERS substrate (Fig. 3(b)). There are corresponding negative shifts in the Ag 3d$_{5/2}$ spectra of the PIERS substrate at 367.5 eV, compared to bulk Ag at 368.3 eV (Fig. 3(d)). The binding energies also indicate that the AgNPs are not oxidised and remain in the Ag(0) metallic state [27]. The opposite-sign shifts in the binding energy of Ag and Ti are attributed to the flow of electrons from TiO$_2$ to AgNPs [20b].

## (d) Electromagnetic and charge transfer interaction between AgNPs and RhB

As we have established a mechanism for charge transfer of photoexcited electrons in TiO$_2$ to AgNPs, there now remain several possibilities for the additional order-of-magnitude Raman signal enhancement of PIERS over SERS. Below, we address each hypothesis and arrive at the conclusion that the only viable explanation is that the strong adsorption of RhB to Ag increases the intrinsic Raman polarizability, and this polarizability is further enhanced by photo-induced charge transfer from TiO$_2$ to Ag. Charge transfer from AgNPs to RhB is possible as the reduction potential E$^0$(RhB/RhB$^+$) [28] lies below the Fermi level of AgNPs shown in Fig 2.

1. Increased electron density on AgNPs increasing the electromagnetic enhancement: As more electrons are transferred to AgNPs, the electromagnetic field around the particles will be enhanced, which can create stronger Raman signals. However, we can discount the purely electromagnetic contribution to the PIERS enhancement due to increased electron density, as this is not predicted to increase the enhancement factor by more than 13% [29]. The theoretical prediction is supported by the experiments by Mulvaney *et. al.* [30] and Dos Santos *et. al.* [31].

2. Suppression of the fluorescence background of RhB due to interaction with AgNPs or photoexcited TiO2: Marchi *et. al.* [32], showed that chemi-sorption of RhB onto AgNPs in solution did not quench the fluorescence of RhB. This is supported by the non-background corrected Raman spectra collected in Fig S7, which shows an increase rather than a decrease in the background fluorescence. We were also unable to measure any Raman enhancement of RhB on bare amorphous or defective $TiO_2$. Hence, the fluorescence quenching mechanism can be discounted in our case.

3. Metal-to-ligand-charge-transfer (MLCT) resonance Raman enhancement: A third possibility, which is favoured by current studies [10a, 10g, 12, 13c, 15], is ligand-to-metal charge transfer (LMCT) or metal-to-ligand charge transfer (MLCT) bands forming between Ag and RhB, which adds a further resonance Raman enhancement [1] in addition to the plasmonic field enhancement. Unfortunately, it is not possible to optically measure the spectrum of the charge transfer band, as it would be obscured by the plasmonic resonance. Hence, the energy level diagram in Fig. 4 and Fig. S15 was rigorously constructed (SI Section VI) using a range of spectroscopic/electrochemical techniques (XPS, UV-Vis, cyclic voltammetry) and computational methods. Despite there being three different measurements [28a, 28c, 33] of the HOMO & LUMO levels of RhB, none allow for resonant excitation by the incident laser wavelength. There is no transition possible from any of the occupied energy levels (oxygen vacancy state, valence band of $TiO_2$, Fermi level of AgNPs and HOMO of RhB) to unoccupied levels (conduction band of $TiO_2$ and LUMO of RhB) resonant with a photon at 488 nm (2.54 eV).

4. Photo-induced electron transfer from to RhB increasing intrinsic Raman polarizability: The final explanation for PIERS enhancement is a change in the intrinsic Raman cross-section of the chemisorbed molecule on the AgNPs' surface v.s. the cross-section of a non-chemisorbed molecule[3]. The binding energies of three different adsorption geometries were calculated in Fig. 4(b): via the amine functional group ($RhB^+$-amine-Ag), the stacking of the xanthene ring on Ag ($RhB^+$-xanthene-Ag), or via the carboxyl group ($RhB^+$-carboxyl-Ag) [34]. The binding via the carboxyl group is very unfavourable, while the binding via the xanthene and amine groups are very close in terms of energetic stability.

5.

The most likely binding geometry was determined by calculating the Raman modes of $RhB^+$ using DFT in Fig 4(a) (B3PW91/6-31G+(d,p) and LANL2DZ(for Ag)), and applying Moscovitz's SERS surface selection rules[35]: normal modes with a large polarizability component normal to the metal's surface will be enhanced. In Fig 4(a), both the PIERS and SERS substrates show an enhanced Raman intensity of the antisymmetric xanthene ring stretch at ~1370 cm$^{-1}$ (theory: 1344 cm$^{-1}$) and symmetric amine stretches at ~1200 cm$^{-1}$ (theory: 1260cm$^{-1}$) relative to the pure xanthene ring stretch at 1646 cm$^{-1}$. As the enhanced xanthene ring stretches have their displacement vectors aligned along the long axis of the RhB molecule (Fig. 4(c)), this implies that the binding needs to occur via a functional group face that is orthogonal to RhB's long axis. This leaves only one option: the nitrogen fragment of the RhB must bind strongly to the Ag surface. This conclusion is supported by the increased intensity of peaks associated with a xanthene ring carbon-nitrogen stretch (1380-1390 cm$^{-1}$). If RhB were lying fully flat, it would be the out-of-plane modes that would be enhanced, at the expense of the in-plane modes – which is not observed. Marchi *et. al.* [32] also found that RhB prefers to coordinate to Ag via its amine groups. As the amine groups motion is hindered, the florescence quantum yield should increase as the greater rigidity reduces the rate of internal conversion – which we observe as an overall increase in the Raman background in Figure S7.

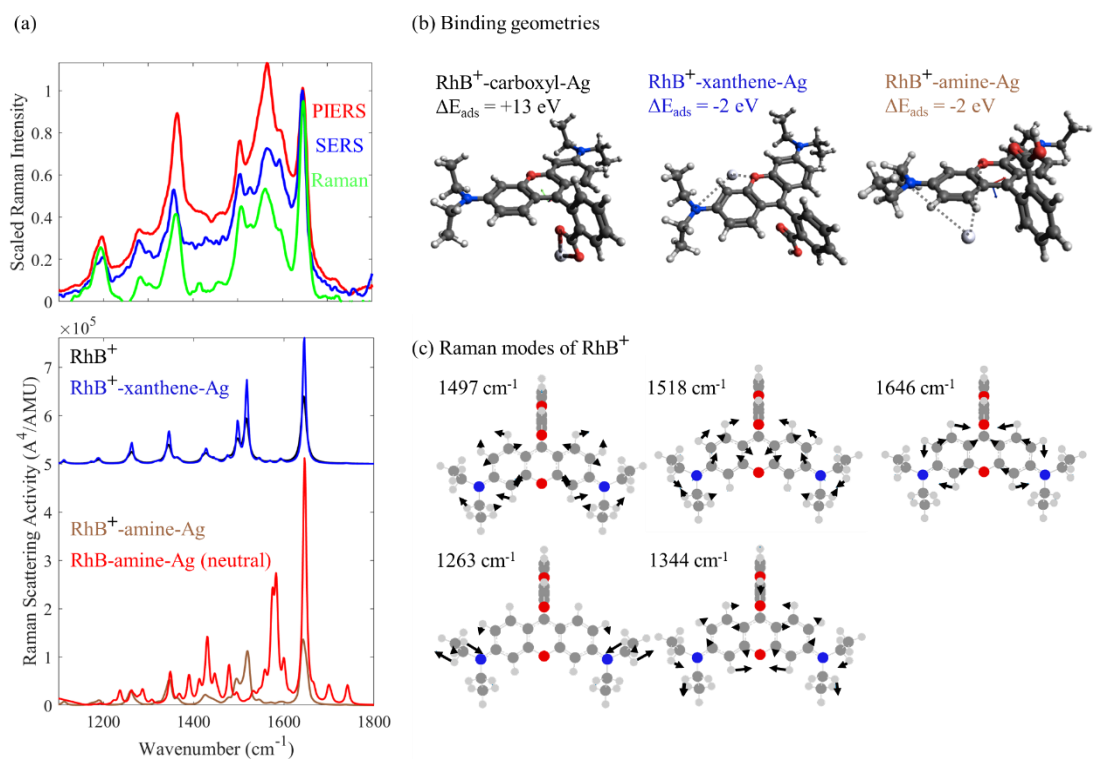

*Figure 4. Comparison of experimental and simulated PIERS, SERS, and normal Raman spectra. (a) Top: Scaled Raman spectra of PIERS, SERS and normal Raman spectra of RhB (scaled to the peak intensity at 1647 cm$^{-1}$). (Bottom:) Dynamic Raman activities predicted with DFT (B3PW91/6-31G(d,p)+ LANL2DZ) at 488 nm excitation for different binding geometries of RhB on Ag and normal RhB cation. Raman transition frequencies were scaled by 0.963 to match that of the experimental band at 1648 cm$^{-1}$. (b) Associated binding geometries of RhB on Ag (bound via carboxyl group, via xanthene ring's oxygen, and via amine group) and associated DFT-calculated adsorption energies (relative to unbound RhB and Ag). (c) Raman-active vibrational modes of RhB cation calculated via DFT. Arrows indicate the*

*magnitude and direction of the displacement vectors for each vibration*

The impact of charge transfer was modeled by calculating the dynamic Raman scattering activity at an excitation wavelength of 488 nm for the most likely binding geometry (RhB$^+$-amine-Ag). The RhB$^+$-amine-Ag spectrum has a very similar spectrum and scattering intensity to the bare RhB+ cation. Upon electron donation to form the neutral RhB-amine-Ag complex, the overall predicted Raman scattering intensity increases, and the band at 1565 cm$^{-1}$ is now enhanced. The combination of the cationic RhB$^+$-amine-Ag and neutral RhB$^+$-amine-Ag predicted spectra now matches the experimental SERS and PIERS spectra very well. The experimental PIERS spectrum has a higher peak intensity at 1565 cm$^{-1}$ as the proportion of neutral molecules is larger than cationic molecules due to efficient charge transfer.

# Conclusions

Photo-Induced Enhanced Raman Spectroscopy (PIERS) was characterised using a heterostructure of defective crystalline TiO$_2$ and AgNPs created using self-organized growth of TiO$_2$ nanotubes and thermal dewetting of a Ag film to create AgNPs. The PIERS substrate showed an order-of-magnitude greater enhancement of Raman signals than the equivalent SERS substrate without defective crystalline TiO$_2$.

We show clear evidence that the key steps in the PIERS mechanism are: (i) absorption of visible light by defect states in TiO$_2$, (ii) population of the conduction band of TiO$_2$, (iii) electron transfer to the Fermi level of AgNPs, and (iv) electron transfer from AgNPs to the adsorbed RhB molecule. The final step increases the intrinsic polarizability of the molecule. Contrary to previous reports [9a, 10h], we find no evidence for charge transfer resonance Raman enhancement due to photon-assisted electron transfer from the substrate to Rhodamine dyes. We also highlight the balance that needs to be struck between defects that increase Raman laser absorption and crystallinity that increases charge mobility and transfer to AgNPs.

By identifying the elementary steps of the PIERS enhancement process, the design rules for fabricating a PIERS substrate are clarified: A PIERS substrate requires a crystalline semiconductor capable of absorbing the visible Raman excitation laser (via sub-bandgap defect states or an appropriate bandgap), a semiconductor conduction band energy greater than the Fermi level of plasmonic nanoparticles, a probe molecule capable of undergoing charge transfer and whose Raman polarizability is enhanced and not suppressed upon charge transfer. These detailed requirements set limits on the range of molecules and semiconductor systems which can exploit the PIERS effect, and raises the question of whether the PIERS effect can be as broadly applied as SERS.

Further experiments that could elucidate more of the details of the individual steps above include ultrafast transient absorption and time-resolved fluorescence measurements to determine various charge transfer rates. We are also investigating the effect of the relative proportion of anatase to rutile phases of $TiO_2$ on the PIERS enhancement, along with the effect of defect concentration on the charge transfer and PIERS enhancements and the use of visible-bandgap semiconductors (e.g. $WO_3$) for PIERS. The exact nature of the defects in $TiO_2$ will strongly affect the energy levels within the bandgap of the semiconductor [36], and the characterization of those will be a next step to fully elucidate the effect of oxygen vacancies on charge transfer.

## Contributions

R.A. and J.Y. performed the experiments, analyzed the data, and wrote the paper. J.D. aided the electrochemical anodization experiments and M.K.N. helped interpret the Raman spectra. All authors contributed to the writing of the paper. M.C.S. and W.G. supervised the project.

## Acknowledgements

We thank Dr. Shanghai Wei for assistance with the TEM data collection, Dr. Colin Doyle for assistance with XPS data collection, and Dr. Balan Zhu for electrochemistry and sputtering experiments. We also thank Prof David Geohegan and Dr. Baptiste Auguie for useful discussions. We acknowledge grant funding from Ministry of Business, Innovation and Employment Grants (MBIE) (UOAX1202, UOAX1416).

**Keywords:** SERS, anodization, PIERS, spectroscopy, plasmonics

**Table of Contents graphic**

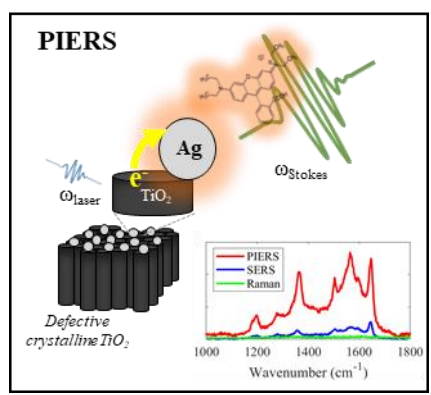

The mechanism of photo-induced enhanced Raman spectroscopy (PIERS), which has an order-of-magnitude higher Raman signal enhancement than typical surface enhanced Raman spectroscopy (SERS), was found with a range of optical and X-ray spectroscopies on a $TiO_2$-

Ag structure. The key source of the PIERS enhancement was an increase in polarizability of the adsorbed probe molecule upon photo-induced charge transfer. A balance between TiO$_2$'s crystallinity and oxygen vacancies was critical to achieve high PIERS enhancements.